# A Hybrid Neural Network with Smart Skip Connections for High-Precision, Low-Latency EMG-Based Hand Gesture Recognition

Hafsa Wazir[1], Jawad Ahmad[2], Muazzam A. Khan[1], Sana Ullah Jan[3], Fadia Ali Khan[4], Muhammad Shahbaz Khan[3]
[1]ICESCO Chair for Big Data Analytics and Edge Computing, Quaid-i-Azam University, Islamabad, Pakistan,
Emails: wazirhafsa@cs.qau.edu.pk, muazzamm.khattak@qau.edu.pk
[2]Cybersecurity Center, Prince Mohammad Bin Fahd University, Alkhobar, Saudi Arabia.,
Email: jahmad@pmu.edu.sa
[3]School of Computing, Engineering and the Built Environment, Edinburgh Napier University, Edinburgh, UK,
Emails: {muhammadshahbaz.khan, s.jan}@napier.ac.uk
[4]Department of Computer Science, HITEC University, Taxila, Pakistan,
Email: fadia.ali@hitecuni.edu.pk

*Abstract*—Electromyography (EMG) is extensively used in key biomedical areas, such as prosthetics, and assistive and interactive technologies. EMG signals measure the electrical activity of muscles during different motions. EMG signals play a key role in gesture recognition studies, such as hand gesture recognition. This paper presents a new hybrid neural network named ConSGruNet for precise and efficient hand gesture recognition. The proposed model comprises convolutional neural networks with smart skip connections in conjunction with a Gated Recurrent Unit (GRU). The proposed model is trained on the complete Ninapro DB1 dataset. The proposed model boasts an accuracy of 99.7% in classifying 53 classes in just 25 milliseconds. In addition to being fast, the proposed model is lightweight with just 3,946 KB in size. The fast inference time and lightweight architecture of the proposed model makes it suitable for resource constrained IoT devices. Moreover, the proposed model has also been evaluated for the reliability parameters, i.e., Cohen's kappa coefficient, Matthew's correlation coefficient, and confidence intervals. The close to ideal results of these parameters validate the models performance on unseen data.

## I. Introduction

Hand gestures play a critical role in day to day human interactions. The detection and classification of hand gestures is important in interpreting hand movements to machines and in assisting intuitive interfaces to control various devices and robots. In addition, it's also important in biomedical applications, such as development of prosthetic devices, rehabilitation, muscle order diagnostics, etc. For this purpose, Electromyography (EMG) classification is performed, where electrical signals generated by muscle activities while performing hand movements, are classified for analysis and interpretation, since raw data can not be used directly due to its nonlinearity and nonstationary characteristics [1]. EMG provides a rich, physiological perspective to gesture recognition, which is important for developing responsive and intuitive systems. This area is a key to progressing in artificial intelligence, and aims to understand complex hand movements and turn them into clear instructions. In essence, EMG classification serves as a bridge between human biological signals and various applications, enhancing the interaction between humans and machines, and plays a pivotal role in medical diagnostics and therapeutic interventions.

Various technologies have been used for EMG classification such as signal processing (time-domain and frequency-domain analysis etc.), feature selections methods (fourier transformation, energy features etc.) [2] [3], machine learning and deep learning techniques (SVM, ANN, CNN, LSTM etc.) [4] [5] [6]. And results from literature show that applying neural networks in gesture recognition by using EMG marks a significant leap in pursuing more intuitive and adaptive systems. In this research, a new hybrid neural network has been designed specifically for classification of EMG signals for hand movement recognition. The hybrid model is a combination of 1 Dimensional Convolutional Neural Network (CNN) using Smart Skip Connections [7], and the Gated Recurrent Unit (GRU) named ConSGruNet.

Incorporating smart skip connections into neural networks addresses the challenges of depth and complexity. These connections allow for selective bypassing of layers within the network, facilitating enhanced feature propagation and mitigating the risk of information loss across layers. This architectural innovation is pivotal in balancing extracting detailed features and understanding high-level gesture semantics. This study explores the potential to create more responsive and accurate systems by integrating EMG data by leveraging hybrid neural network ConSGruNet with smart skip connections for hand gesture recognition using EMG. Furthermore, this research pushes the boundaries of machine learning and AI, contributing to our understanding of how complex physiological data can be effectively integrated into computational models. Figure 1 illustrates the core methodology adopted for the classification of hand movements using EMG signals.

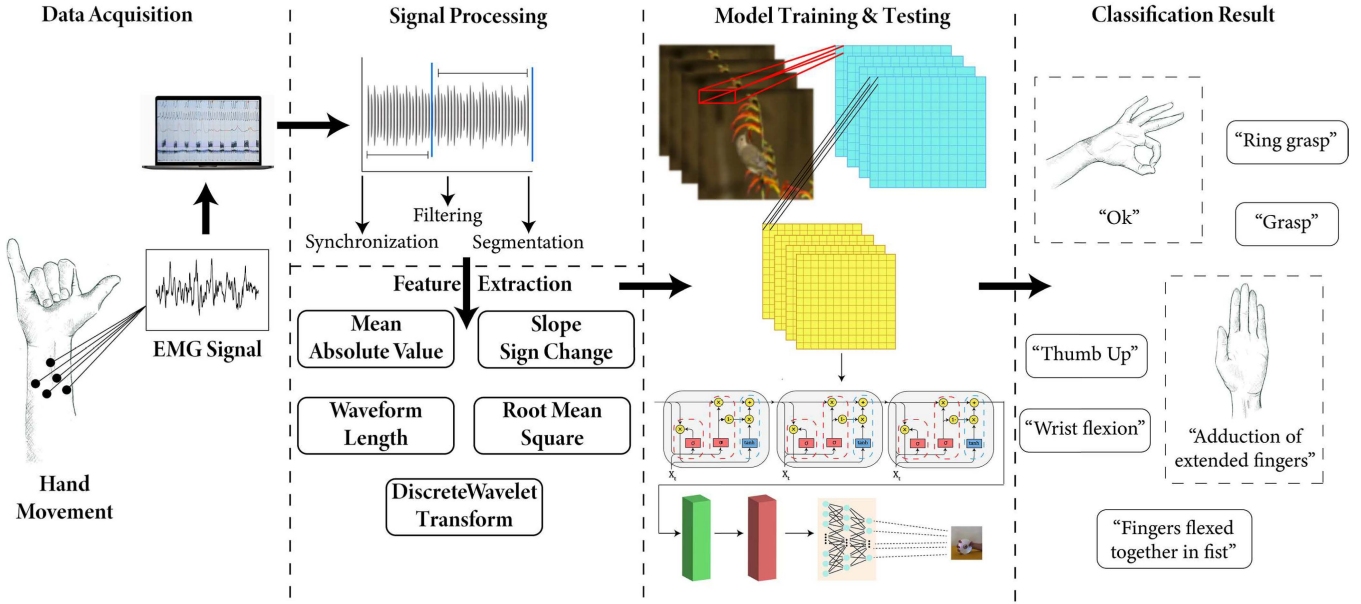

Fig. 1: The Adopted Methodology for EMG based Hand Movement Classification

The main contributions of this research are as follows:

1) Developed a unique one dimensional Convolutional Gated Recurrent Unit using Smart Skip connections - ConSGruNet specifically tailored for hand movement classification using EMG, introducing an innovative model that enhances accuracy significantly.
2) Developed a highly accurate and resource-efficient model for IoT devices, achieving a 25-millisecond computation time with only 985,973 parameters, and a compact size of 3,946 KB.
3) Utilized the Ninapro DB1 dataset consisting of 10 channels, encompassing the extensive range of classes and individuals, classifying 53 classes from 27 individuals with the highest classification accuracy of 99.7% to date (to the best of the authors' knowledge).

The following sections of the papers are as follows: Section II presents the related literature, section III explains the proposed model and its architectural intricacies, section IV expands on the dataset and evaluation criteria used in this study, section V presents the performance results, section VI details the comparative analysis performed for the proposed model, and finally section VII concludes the paper.

## II. LITERATURE REVIEW

Several papers can be found in literature, which intended to optimise the hand gesture recognition, for instance, Muhammad Farrukh Qureshi et al. [8] presents E2CNN, a novel CNN for classifying sEMG signals. This efficient concatenated CNN is optimized for processing Log-Mel (LM) spectrogram images of sEMG signals. The model has been tested on two datasets: a longitudinal dataset of ten able-bodied and six trans-radial amputee subjects and 10 subjects from the publicly available NinaPro DB1 dataset. E2CNN demonstrated high-performance accuracy (98.31% ± 0.5% for able-bodied and 97.97% ± 1.41% for amputee subjects; 91.27% on NinaPro DB1), outperforming baseline CNN models. Similarly, Xu et al. [9] utilised surface electromyography (sEMG) signals for hand gesture recognition. The authors developed a Concatenate Feature Fusion (CFF) strategy to enhance the performance of Recurrent Convolutional Neural Networks (RCNNs). The CFF-RCNN model demonstrates classification accuracies of 88.87% on DB1, 99.51% on DB2, and 99.29% on DB4 from the NinaPro database. The study highlights the effectiveness of the CFF strategy in improving model performance.

Furthermore, Karnam et al. [10] presented a new method for classifying hand movements. The authors presented EMGHandNet, a hybrid model combining CNNs and Bidirectional Long Short-Term Memory (Bi-LSTM) networks. The paper reported improved classification for various datasets, such as NinaPro DB1, DB2, DB4, and UCI Gesture. Besides, Yuhuang Zheng [11] presented a novel framework for classifying hand movements using data glove technology. The study introduces a classification system comprising two parts: a movement detection algorithm centred around the Fine K-nearest neighbour (Fine KNN) method and a movement classification algorithm that is a combination of downsampling in data preparation and a new deep learning network named the DBDF network, primarily based on Bidirectional Long Short-Term Memory (BiLSTM). This framework was tested on the Ninapro DB1 dataset, demonstrating the ability to classify a broader range of hand movements (52 types) with high accuracy of 93.15%, using only one data glove.

Similarly, Reza Bagherian Azhiri et al. [12] utilised recurrent neural networks (RNNs) for real-time classification of sequential EMG signals, focusing on improving classification accuracy and reducing delay time. The study compares various RNN-based architectures using discrete wavelet transform for feature extraction, achieving notable results. Specifically, the Bidirectional Recurrent Neural Network (BRNN) with the same inputs outperforms other architectures, reaching 96% classification accuracy in just 600 milliseconds for the two-channel (2C) dataset. In addition, Pancholi et al. [13] proposed an advanced Convolutional Neural Network (CNN) framework specifically designed for pattern recognition in surface electromyography (sEMG) signals for upper limb prosthesis control. The framework is evaluated using five benchmark datasets, three of them being Ninapro DB1, DB2 and DB3. It demonstrated its ability to accurately classify various upper limb activities, notably achieving 91.11% accuracy for a dataset with 53 classes of DB1, 89.45% for 49 classes of DB2, 81.67% for 49 classes of DB3 and 95.67% for 6 classes of a dataset with force variation, and 99.11% for 8 classes of a dataset with arm position variation.

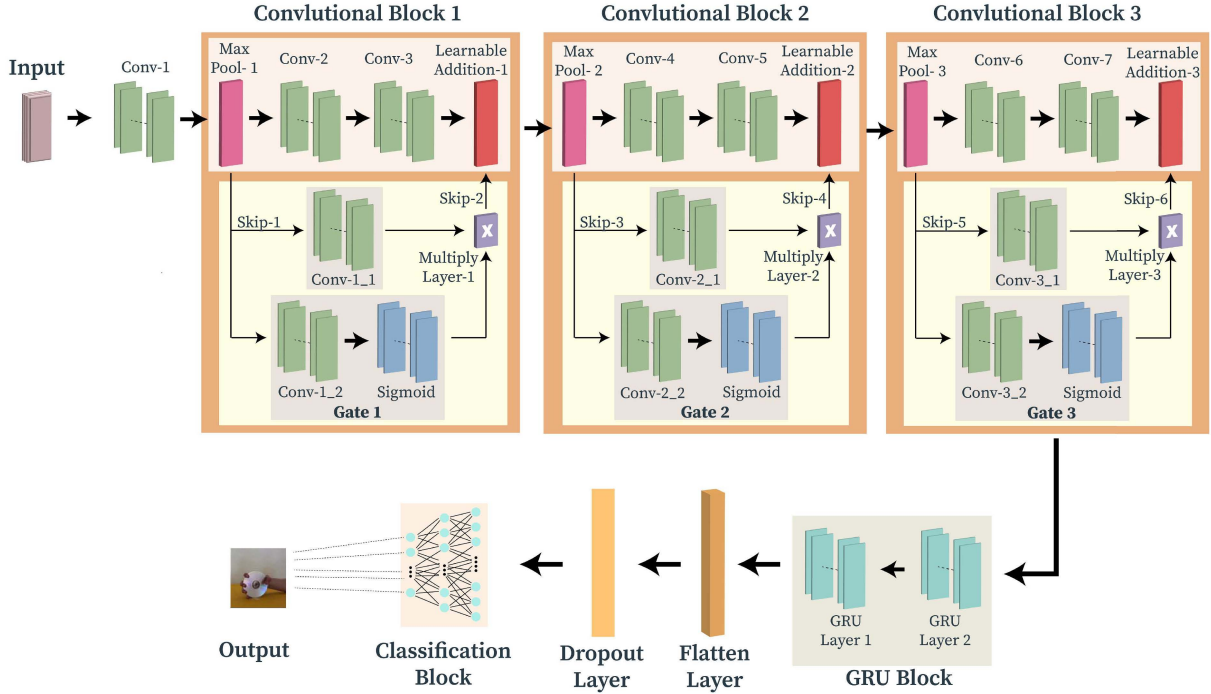

Fig. 2: The Proposed Model: ConSGruNet

The literature review showcases significant advancements in hand gesture classification using electromyography (EMG) signals, employing various machine learning and deep learning methodologies. Despite that, certain gaps and drawbacks remain evident. Many studies, while showing high accuracy, are tested on limited data, raising questions about their generalizability across diverse real-world scenarios. Additionally, the focus on specific EMG features and machine learning models in some studies may not fully capture the complexity of hand movements. Furthermore, while some models boast high accuracy, the computational efficiency and real-time application in prosthetic control still require further exploration and improvement. This study aims to fill the research gaps present in the literature by using large number of hand movements from a wide variety of subjects to ensure generalizability of the model while making sure that it is light-weight, easy to deploy on resource deficient devices and provides a high accuracy in the short time.

## III. THE PROPOSED MODEL CONSGRUNET

This research presents a hybrid neural network ConSGruNet, a combination of one dimensional CNN having smart skip connections and GRU layers. The uniqueness of this model is due to the smart skip connections having gated mechanisms, enabling dynamic information flow management. Figure 2 illustrates the architecture of ConSGruNet.

## IV. DATASET & EVALUATION CITERIA

The dataset used in this study is the Ninapro DB1 dataset, it is a comprehensive resource in electromyography (EMG) research, particularly for gesture and motion classification [14] [15]. It is a publicly available benchmark dataset (ninapro.hevs.ch). It features surface EMG data recorded from 27 subjects performing a variety of hand and wrist movements. This dataset includes 53 exercises (including rest), encompassing basic movements and more complex gestures. Each subject repeated the hand movements displayed on the screen 10 times with a rest of 3 seconds in between. It is an invaluable tool for developing and testing EMG-based classification algorithms, offering a rich set of data points for robust analysis.

The evaluation criteria for this research was divided into two main categories such as performance and reliability. The metrics used for performance are: accuracy, precision, recall, the F1 score calculated through confusion matrix shown in Fig. 4. Meanwhile, Cohen's Kappa, Mathew's Correlation Coefficient, and 95% confidence interval are used to check reliability.

## V. RESULTS & PERFORMANCE EVALUATION

In this section, the performance and reliability of the proposed model will be analysed using the evaluation criteria mentioned in section IV. The performance results are shown in table I. The results show that the model is able to accurately classify all the hand movements.

And the results for the reliability metrics are given below:

- Cohen's kappa is a statistical measure that accounts for the possibility of agreement occurring by chance. It is beneficial when evaluating the reliability of a classification model by considering the agreement between predicted and actual labels.

$$K = \frac{P_e - P_o}{1 - P_e} \quad (1)$$

The given equation 1 represents Cohen's Kappa, where $P_e$ is the probability of expected agreement of the annotators on the randomly assigned labels - i.e the agreement between the actual class values and model predictions, and $P_o$ is the probability of agreement on the assigned label, essentially the model's overall accuracy. The Cohen's Kappa value for the classes was 0.99 on average as shown in Fig. 5.

- The Matthews correlation coefficient is a measure that considers true positives, true negatives, false positives, and false negatives. It is especially effective in handling imbalanced datasets, providing a balanced measure even when class distribution is skewed. The value of Matthew's correlation Coefficient for the classes was 99.8 on average as shwon in Fig. 5. Following equation 2 shows the equation for calculating the Matthew's correlation coefficient.

$$MCC = \frac{TP \times TN - FN \times FP}{\sqrt{TP + FP + TN + FN}} \quad (2)$$

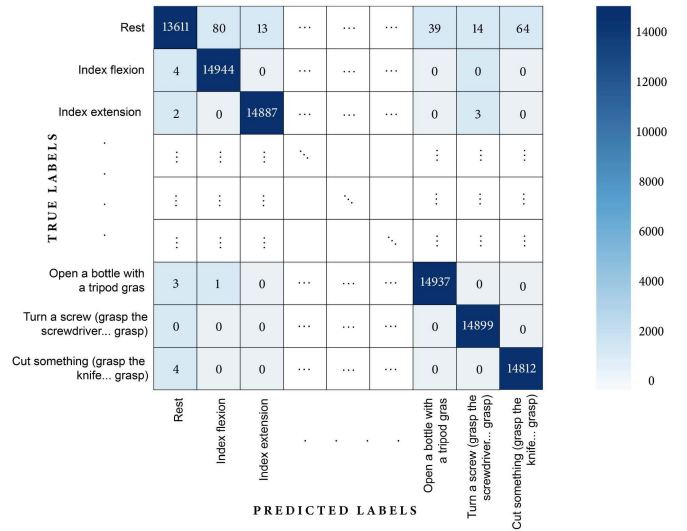

Fig. 4: Confusion matrix depicting the performance of the proposed model

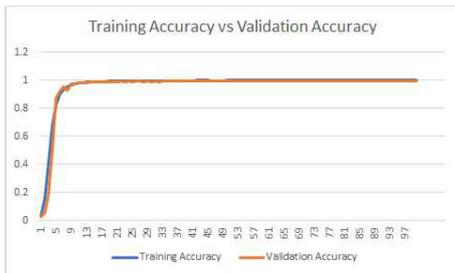

(a) Accuracy curves

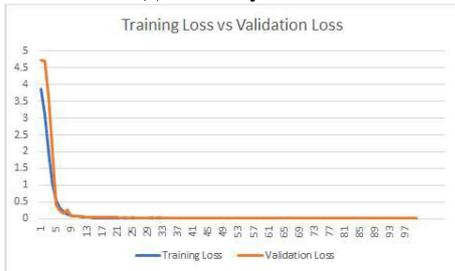

(b) Loss curves

Fig. 3: Training performance of ConSGruNet.

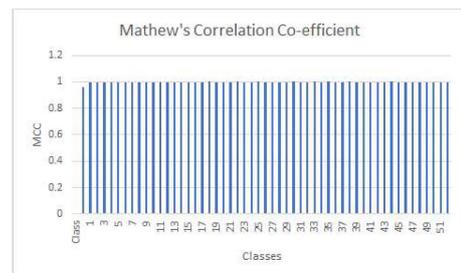

(a) Matthew's correlation coefficient

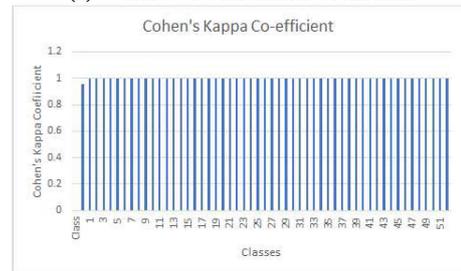

(b) Cohen's kappa coefficient

Fig. 5: Reliability parameters.

TABLE I: Class-Wise Classification Report Using Performance Metrics

| Class | Precision | Recall | F-1 Score | Class | Precision | Recall | F-1 Score |
|---|---|---|---|---|---|---|---|
| Rest | 1.00 | 0.92 | 0.96 | Wrist radial deviation | 1.00 | 1.00 | 1.00 |
| Index flexion | 0.99 | 1.00 | 1.00 | Wrist ulnar deviation | 1.00 | 1.00 | 1.00 |
| Index extension | 1.00 | 1.00 | 1.00 | Wrist extension with closed hand | 1.00 | 1.00 | 1.00 |
| Middle flexion | 1.00 | 1.00 | 1.00 | | | | |
| Middle extension | 1.00 | 1.00 | 1.00 | Small diameter grasp (power grip) | 1.00 | 1.00 | 1.00 |
| Ring flexion | 1.00 | 1.00 | 1.00 | Fixed hook grasp | 1.00 | 1.00 | 1.00 |
| Ring extension | 1.00 | 1.00 | 1.00 | Index finger extension grasp | 1.00 | 1.00 | 1.00 |
| Little finger flexion | 1.00 | 1.00 | 1.00 | Medium wrap | 1.00 | 1.00 | 1.00 |
| Little finger extension | 1.00 | 1.00 | 1.00 | Ring grasp | 1.00 | 1.00 | 1.00 |
| Thumb adduction | 1.00 | 1.00 | 1.00 | Prismatic four fingers grasp | 1.00 | 1.00 | 1.00 |
| Thumb abduction | 1.00 | 1.00 | 1.00 | Stick grasp | 1.00 | 1.00 | 1.00 |
| Thumb flexion | 1.00 | 1.00 | 1.00 | Writing tripod grasp | 1.00 | 1.00 | 1.00 |
| Thumb extension | 1.00 | 1.00 | 1.00 | Power sphere grasp | 1.00 | 1.00 | 1.00 |
| Thumb up | 1.00 | 1.00 | 1.00 | Three finger sphere grasp | 1.00 | 1.00 | 1.00 |
| Extension of index and middle, flexion of the others | 1.00 | 1.00 | 1.00 | Precision sphere grasp | 1.00 | 1.00 | 1.00 |
| Flexion of ring and little finger, extension of the others | 1.00 | 1.00 | 1.00 | Tripod grasp | 1.00 | 1.00 | 1.00 |
| Thumb opposing base of little finger | 1.00 | 1.00 | 1.00 | Prismatic pinch grasp | 1.00 | 1.00 | 1.00 |
| Abduction of all fingers | 1.00 | 1.00 | 1.00 | Tip pinch grasp | 1.00 | 1.00 | 1.00 |
| Fingers flexed together in fist | 1.00 | 1.00 | 1.00 | Quadpod grasp | 1.00 | 1.00 | 1.00 |
| Pointing index | 0.99 | 1.00 | 1.00 | Lateral grasp | 1.00 | 1.00 | 1.00 |
| Adduction of extended fingers | 1.00 | 1.00 | 1.00 | Parallel extension grasp | 1.00 | 1.00 | 1.00 |
| Wrist supination (axis: middle finger) | 1.00 | 1.00 | 1.00 | Extension type grasp | 1.00 | 1.00 | 1.00 |
| Wrist pronation (axis: middle finger) | 1.00 | 1.00 | 1.00 | Power disk grasp | 1.00 | 1.00 | 1.00 |
| Wrist supination (axis: little finger) | 1.00 | 1.00 | 1.00 | Open a bottle with a tripod grasp | 1.00 | 1.00 | 1.00 |
| Wrist pronation (axis: little finger) | 1.00 | 1.00 | 1.00 | Turn a screw (grasp the screwdriver with a stick grasp) | 1.00 | 1.00 | 1.00 |
| Wrist flexion | 1.00 | 1.00 | 1.00 | Cut something (grasp the knife with an index finger extension grasp) | 1.00 | 1.00 | 1.00 |
| Wrist extension | 1.00 | 1.00 | 1.00 | | | | |

- Confidence intervals are employed to quantify the uncertainty associated with our performance metrics. By providing a range of values within which the model's true performance is likely to lie, confidence intervals contribute to a more comprehensive understanding of the model's stability and generalizability. For the proposed model, the 95% confidence interval was between the range of 0.9974 - 0.9976.

$$ConfidenceInterval = x \pm z\frac{s}{\sqrt{n}} \quad (3)$$

Equation 3 is used to calculate the confidence interval. Where s is the standard deviation, n is the sample size, z is the z-value for the confidence level, x is the mean value, and plus-minus ($\pm$) represents the lower and upper limits.

- Inference time is the period a model takes to process input data and produce a classification. For EMG classification, the preferred range of time is between 0-200 ms, and shorter inference times are desirable as they prevent the delays in mind-to-hand movement. The proposed model achieves a total inference time of only 25 milliseconds. This rapid processing capability makes the model highly suitable for EMG classification tasks, particularly in scenarios requiring immediate response and analysis.

## VI. COMPARATIVE ANALYSIS

To further gauge the proposed model's extensibility, it was compared with the state-of-the-art models currently available in literature. Table II shows the summary of the results from the comparison between proposed model and models from literature.

The results showed that the proposed model performs significantly better from previous models in terms of efficiency, accuracy and adaptability. The comparison showed that the increase in number of classes leads to a decrease in the accuracy of the model. The types of features selected for the EMG classification also affects the accuracy of the model.

## VII. CONCLUSION

This paper presented a novel hybrid neural network with smart skip connections for the accurate and efficient detection of hand movements. The proposed model is intended to be deployed on resource constrained IoT devices as it has a lightweight architecture with only 3,946 Kb size. The proposed model has been trained on the Ninapro DB1 dataset, comprising of 53 classes from 27 individuals and it boasts a remarkable accuracy of 99.7% with a fast inference time of just 25 milliseconds. To gauge the model's generalisability and its performance on unseen data, the proposed model has also been evaluated for the reliability parameters, i.e., Cohen's kappa coefficient, Matthew's correlation coefficient, and confidence intervals. The close to ideal results of these parameters validate the models performance on unseen data and make it suitable to be used in resource constrained IoT devices.

TABLE II: Comparison of Proposed Model with Other Deep Learning Models

| Works | Models | Datasets | Features | Classes | Accuracy | Prediction Time (ms) |
|---|---|---|---|---|---|---|
| Yuhuang Zheng [11] | DBDF | Ninapro DB1 | Internally extracted fused | 52 | 93.15% | × |
| Xu et al. [9] | CFF-RCNN | Ninapro DB1 | MAV, ZC, SSC, WL | 50+ | 88.87% | × |
| Qureshi et al. [8] | E2CNN | Ninapro DB1 | Log-Mel Spectogram Images | 9 | 91.27% | 64.15 ($\pm$) 5.914 |
| Karnam et al. [10] | CNN-BiLSTM | Ninapro DB1 | Energy-Features | 53 | 95.77% | 0.0018 - 0.518 |
| Azhiri et al. [12] | RNN | Dataset by Center Of Intelligent Mechatronic Systems at the University of technology at Sydney | Discrete wavelet decomposition (DWD) | 25 | 96% | × |
| Pancholi et al. [13] | DFPL | Ninapro DB1 | TD, TFD | 53 | 91% | × |
| Wei et al. [16] | MV-CNN | Ninapro DB1 | Low-level & high-level features through CNN | 52 | 88.2% | × |